\documentclass[
    a4paper, 
    amsfonts, 
    amssymb, 
    amsmath, 
    reprint, 
    showkeys, 
    footinbib, 
    longbibliography, 
    twoside,
    notitlepage,
    aps,
]{revtex4-2}

\usepackage[warn]{mathtext}
\usepackage[utf8]{inputenc}

\usepackage{amsthm}
\usepackage{mathtools}
\usepackage{physics}
\usepackage{xcolor}
\usepackage[left=23mm,right=13mm,top=35mm,columnsep=15pt]{geometry} 
\usepackage{adjustbox}
\usepackage{placeins}
\usepackage{csquotes}
\usepackage{float}

\usepackage{blindtext}

\usepackage{bm}

\usepackage{xcolor}

\definecolor{urlcolor}{HTML}{3333B2}
\definecolor{citecolor}{HTML}{3333B2}
\definecolor{black}{HTML}{000000}

\usepackage{xcolor}

\usepackage{graphicx}

\graphicspath{}
\DeclareGraphicsExtensions{.pdf,.png,.jpg}

\usepackage{array}
\usepackage{placeins}
\usepackage{tcolorbox}

\usepackage[english]{babel}
\usepackage{comment}
\usepackage[pdftex, pdftitle={Article}, pdfauthor={Author}]{hyperref} 
\hypersetup{citecolor=citecolor, urlcolor=urlcolor, colorlinks=true}
\let\vec=\mathbf

\begin{document}
 
\title{Acoustic resonators: symmetry classification and multipolar content of the eigenmodes.}
\author{Mariia Tsimokha}
\author{Vladimir Igoshin}
\author{Anastasiya Nikitina}
\author{Mihail Petrov}
\author{Ivan Toftul}
\email{itoftul@itmo.ru}
\author{Kristina Frizyuk}
\email{k.frizyuk@metalab.ifmo.ru}
\affiliation{ The School of Physics and Engineering, ITMO University}

\begin{abstract}
  
    Acoustics recently became a versatile platform for discovering novel physical effects and concepts at a relatively simple technological level. On this way, single resonators and the structure of their resonant modes play a central role and define the properties of complex acoustic systems such as acoustic metamaterials, phononic crystals, and topological structures. In this paper, we present a powerful method allowing a qualitative analysis of eigenmodes of resonators in the linear monochromatic acoustic domain based on multipole classification of eigenmodes. Using the apparatus of group theory, we explain and predict the structure of the scattered field knowing only the symmetry group of the resonator by connecting the multipolar content of incident and scattered fields. Such an approach can be utilized for developing resonators with predesigned properties avoiding time-consuming simulations. We have performed full multipole symmetry classification for a number of resonators geometries, and tightened it with scattering spectra profiles.   
    
    \end{abstract}

\maketitle

\section{Introduction} \label{sec:intro}
Studying acoustic resonators is essential both for many technological application and for fundamental research developing   acoustic metamaterials with established properties~\cite{Ma_Sheng_2016,Guo_Zhang_Fang_Fattah_2018,Cummer_Christensen_Alu_2016,PhysRevLett.127.084301}, various opto-mechanical systems~\cite{Dostart_Liu_Popovic_2017,PhysRevLett.123.183901},topological insulators~\cite{Ni_Li_Weiner_Alu_Khanikaev_2020, Ni_Weiner_Alu_Khanikaev_2019}, and achieving bound states in the continuum~\cite{Pilipchuk_2020,MAKSIMOV201552}. One of the most important characteristics of any resonator is its eigenmode spectrum and their filed structure. Commonly, this problems is addressed with full-wave numerical simulations, while analytical solutions can be defined solely for a limited number of resonator shapes. A spherical scatterer is one of the examples and the plane wave scattering on an arbitrary size sphere~\cite{doi:10.1063/1.5058149,doi:10.1121/1.1906621} was considered for the first time more than 150 years ago by Clebsch~\cite{Clebsch1863Jan} and Lorentz~\cite{lorenz1890lysbevgelser} for elastic waves, which in electro-magnetic theory is well-known as Mie-scattering~\cite{Mie}.  However, the unified  description of eigenmodes  in an acoustic resonator of arbitrary shape has not been made so far. In solid-state physics, quantum chemistry, and optics a poweful method based on group theory analysis has been widely utilized~\cite{landau,Ivchenko_Pikus_1995,Dresselhaus2008,koster1963properties,Hayami2018-Classificationofato}.   The mode structure is defined solely by the symmetry the system and can be  classified by the irreducible representations (irreps) of the system's symmetry group. The  symmetry of the eigenmodes can also give an answer on which modes are involved in physical processes such as linear and nonlinear wave scattering, also referred to as selection rules \cite{Cammarata2019-xn--Phononphonon-p19fscatter,selrul, Frizyuk2019-Second-harmonicgener, PhysRevB.99.075425,  PhysRevLett.120.087402, PhysRevB.102.035434,PhysRevB.103.184302}.

\begin{figure}[ht!]
    \centering
    \includegraphics[width=0.49\textwidth]{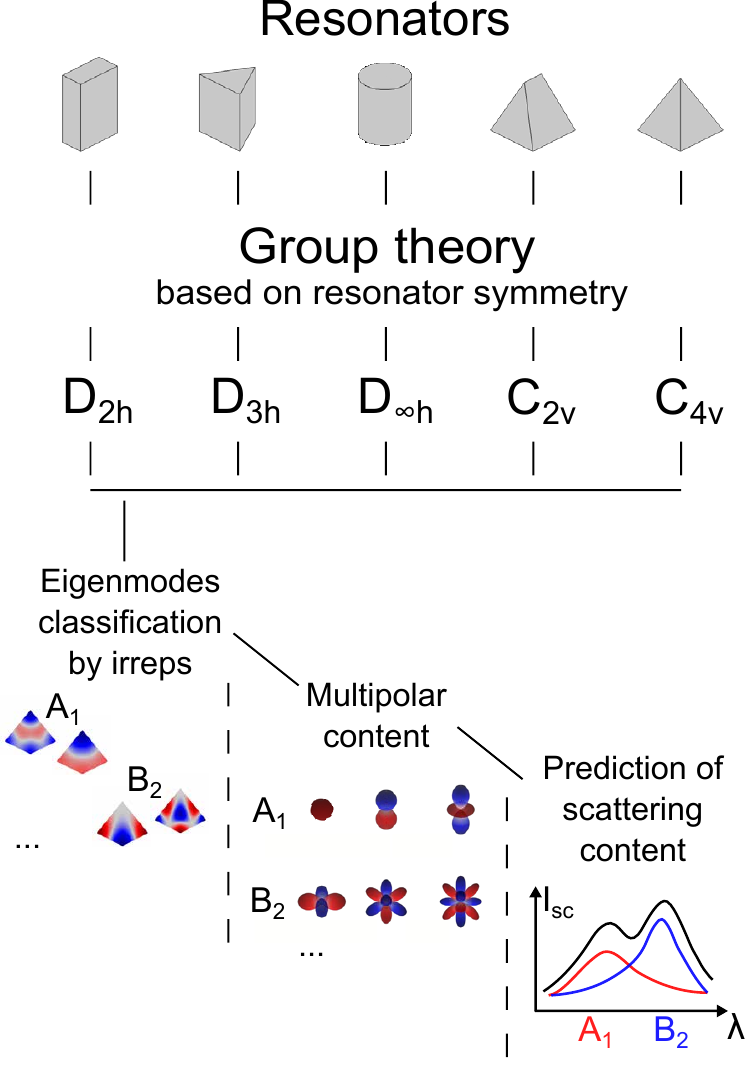}
    \caption{The general approach and step-by-step algorithm to analysis of acoustic resonators modes suggested in this work.}
    \label{concept}
\end{figure}
We build our approach on  multipole decomposition of acoustic waves. Generally, the multipole  expansion is actively used in nanophotonics~\cite{Smirnova_Kivshar_2016,Kivshar_Miroshnichenko_2017,doi:10.1021/acsphotonics.7b01038, krasikov2020multipolar} and demonstrated that it can be effectively used for predicting the optical properties of subwavelength resonators. In acoustics the multipoles approach is not that widely spread however 
started to draw attention recently \cite{Bolton1992, Meng2019, Liu2020}, enabling  effective control over the wave propagation directions and radiation reaction forces~\cite{Toftul2020Dec,Lima2021Jul,Rehfeld_1978, Wei_Rodriguez-Fortuno_2020,Lu_Ding_Wang_Peng_Cui_Liu_Liu_2017,PhysRevLett.123.183901,Gong2021-Equivalencebetweena}.  

In our paper, we provide classification and multipole expansion of an acoustic resonators' eigenmodes of various symmetry, as well as its application to the acoustic scattering.  Throughout this work we will discuss analyze only  longitudinal  acoustic pressure waves in monochromatic domain.  Inspired by the recent progress in  nanophotonics, we will operate in terms of spherical harmonics (multipoles) basis analyzing their symmetry~\cite{doi:10.1143/JPSJ.65.2670, gladyshev,Xiong:20, paper}, which could be even simpler and efficient due to scalar origin of  fields. We show how to connect the symmetry of the resonator with the particular multipolar components of the eigenmode representation. Basing on this, one can immediately interpret and predict the scattering spectra, directivity  of the scattering, and even acoustics forces acting on resonators due to interaction with an arbitrary incident wave.


The manuscript is constructed as follows: in Section~\ref{sec:intro} we give introductory part, which settles up the place of this manuscript in the current state of linear acoustics and optics; in Section~\ref{sec:basic_group} we give some helpful basis of group theory which is necessary for the understanding of the main results; in Section~\ref{sec:results} we explain results obtained for multipole expansion for resonators of $D_{3h}$ symmetry group, and elaborate it on resonators of decreased symmetry; in Section~\ref{sec:scattering} we discuss an influence of resonators' symmetry on the cross section of scattered wave. Finally, in Section~\ref{sec:conclusion} we compare it with the case of optical resonators, whilst demonstrating some similarities and dissimilarities, eventually revealing ways of better understanding both types of scattering and other processes. 

\section{{The basics of group symmetry and acoustic modes analysis} }
\label{sec:basic_group}

For better understanding of the theory the brief summary of several topics is presented below. Here we will introduce the concepts of irreducible representation, functions transformed under irreducible representation (basis of the representation), spherical harmonics, multipole expansion and Wigner theorem.

\begin{figure}[t]
    \centering
    \includegraphics[width = 0.5\textwidth]{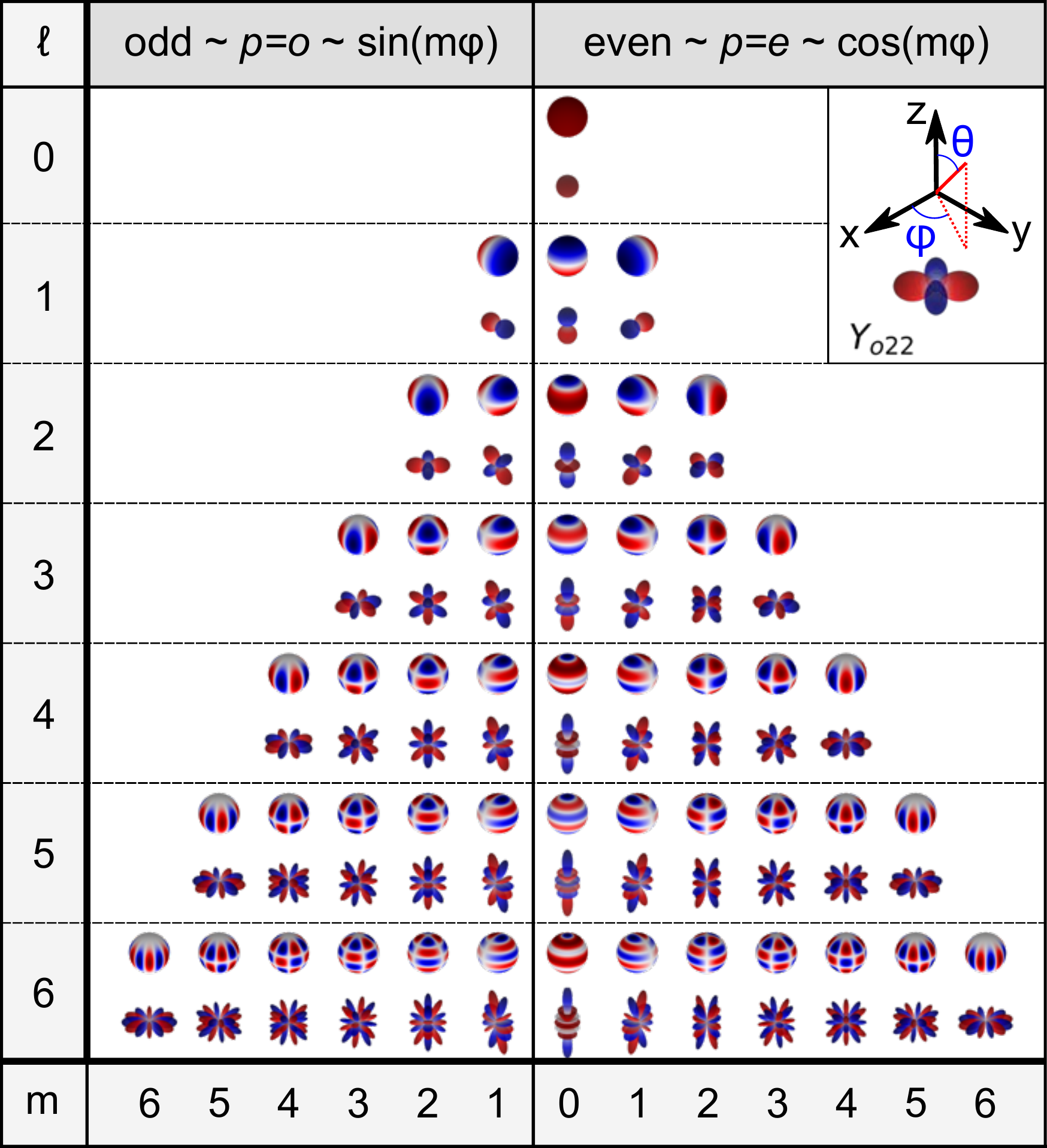}
    \caption{The real spherical functions $Y_{p\ell m}(\theta, \varphi)$ up to $\ell = 6$. At the top of the line --- the color shows the value of the function depending on the angles $\theta, \varphi$, the graph is shown on the sphere. At the bottom --- the radius of the sphere is deformed in proportion to the modulus of the function value.}
    \label{SF_table}
\end{figure}

\begin{figure}
    \centering
    \includegraphics[width=0.37\textwidth]{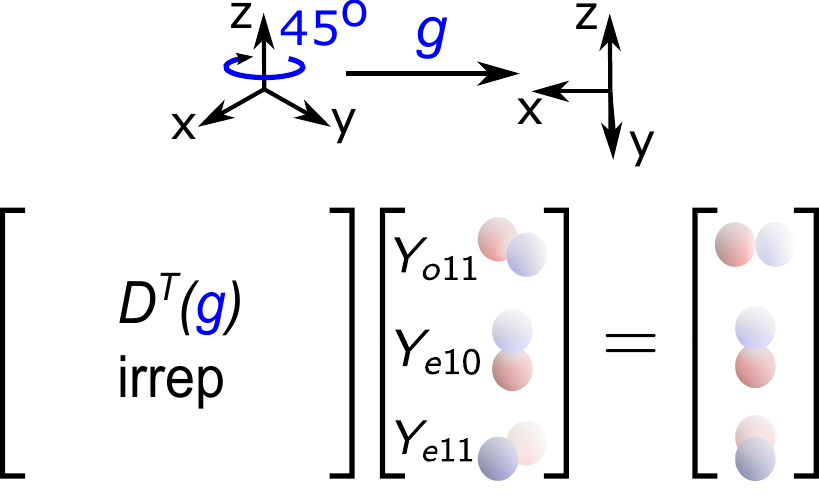}
    \caption{Spherical functions with $\ell=1$ transforming through each other under the same irreducible representation. For complex form of spherical harmonics $D(g)$ matrix can be obtained using Wigner D-matrixes~\cite{SphericalHarmonics_2018}.}
    \label{fig:transofm_ex}
\end{figure}

\subsection{Group and representation theory}

Group is a set equipped with a binary operation that holds three axioms: associativity, identity, and invertibility. The following study focuses mainly on the systems' symmetry groups that consist of the elements that transform the system to itself, i.e. symmetry operations~\cite{BibEntry2001Sep,BibEntry2021Mar}.
A representation of a group $\mathbf{G}$ on a vector space $\mathbb{V}$ is a homomorphism $\mathrm{T}$ of $\mathbf{G}$ to the group of automorphisms  of $\mathbb V$: $\mathrm{GL}(\mathbb{V})$~\cite{Fulton2004-RepresentationTheory}. 
$$\mathrm{T}: \mathbf{G} \rightarrow \mathrm{GL}(\mathbb{V})$$
In simpler words, group representation is a matrix group with box matrixes,  where we assign a matrix $D(g)$ to each element $g \in G$ such that $D(g_1g_2)=D(g_1)D(g_2)$, i.e. the matrixes satisfy the group's multiplication table~\cite{Dresselhaus2008}.
Representation is considered irreducible if there is no any nontrivial invariant subspace in space 
$\mathbb{V}$~\cite{Ivchenko_Pikus_1995,Dresselhaus2008}.
In other words, all matrixes $D(g)$ of any representation  can be simultaneously reduced (by a linear basis transformation) to a block-diagonal form, which consists of irreducible blocks, that turn out to be in fact irreducible representations of that group~\cite{Dresselhaus2008}. 

The term {\it{set of functions $\psi_{i}(\mathbf{r})$ transforming through each other under the irreducible representation}} (or the basis of representation) describes that after the group element action (rotations or reflections in out case) functions are transformed to the particular linear combinations of themselves. 
\begin{equation}
\psi_{i}\left(g^{-1} \mathbf{r}\right)=\sum_{j} D_{j i}(g) \psi_{j}(\mathbf{r})
\end{equation}
To illustrate it, let us introduce spherical harmonic functions also referred as multipoles. A real form in terms of complex spherical harmonics (Fig.  \ref{SF_table}) is set as~\cite{VMK,Rehfeld_1978}:
\begin{equation}
Y_{p\ell m} = 
 \begin{cases}
    \displaystyle\frac{i}{\sqrt{2}}\left(Y^m_\ell - (-1)^m Y^{-m}_\ell\right) & \text{ }\ p=o,\\
    Y_l^0 & \text{ }\ m=0,\\
    \displaystyle\frac{i}{\sqrt{2}}\left(Y^{-m}_\ell + (-1)^m Y^{m}_\ell\right) & \text{ }\ p=e.
 \end{cases}
 \label{sh_comlex_to_real}
\end{equation}

Spherical functions with a particular $\ell$ are basis functions of $2\ell+1$-dimensional irreducible representation of the rotation group of sphere $SO(3)$, therefore {\it{under an arbitrary angle rotation they transform into the linear combination of functions with  the same  $\ell$}} ( Fig. \ref{fig:transofm_ex})~\cite{SphericalHarmonics_2018}.

The results of our work are based on a Wigner's theorem \cite{Piroth2007-FundamentalsoftheP}. It is formulated as follows: 

	\begin{equation}
		\mathcal{H}(\mathbf{r})\psi(\mathbf{r})=\epsilon\psi(\mathbf{r})
		\label{eq:theorem_wigner}
	\end{equation}
	{\it{Suppose that an eigenvalue equation, describing a system is invariant under the transformations of a symmetry group, then the eigenfunctions are transformed under irreducible representations of the group.}}
	
Applied to acoustic waves among medium characterized by compressibility $\beta(r)$ and density $\rho(r)$ Helmholtz equation is set as:
\begin{equation}
    - c(r)^2 \nabla^2 p = \omega^2 p, \; c(r)^2 = \displaystyle\frac{1}{\beta(r) \rho(r)}
\end{equation}
here $p$ is the  pressure function, $c(r)$ is the coordinate dependent speed of sound, and the operator $\mathcal{H}(\mathbf{r})\equiv-c(r)^2 \nabla^2$. At first, We  restrict ourselves to considering the hard boundary condition, $\partial_n p|_{\partial \Omega}=0$, at the resonator surface. Alternatively, the radiative, Sommerfeld-type,  boundary condition at $r\to\infty$ can be considered, however from the symmetry point of view they will provide the same result.   
 

From the Wigner's theorem it immediately follows that  the degree of degeneracy of an energy level equals to the dimension of the corresponding irreducible representation.  One of the simplest examples of this approach, is that the each mode of a spherical resonator can be portrayed as a particular spherical function  (Fig.\ref{SF_table})\cite{Dresselhaus2008}, while modes with identical $\ell$  are $(2\ell+1)$-degenerate.

\subsection{Multipole expansion of resonators' eigenmodes}

While the eigenmodes of a spherical resonator are defined by only one spherical harmonic, the situation becomes much more complex for the  resonators of an arbitrary shape. Now, their modes cannot be defined by a specific spherical harmonic but rather can be decomposed over a multipoles set. At this stage, defining the multipole content becomes a complex numerical problem, however, one immediately identify it using the \textit{Wigner's theorem}. Indeed, the modes' behavior under all symmetry transformations defines under which irreducible representation it transforms. Now, the  multipoles contained in the mode should only belong to the same irreducible representation.  Thus, one needs to know, under which irrep of the resonator's symmetry group each spherical harmonics transforms.

\section{Theory and results of multipole expansion} 
\label{sec:results}
\subsection{Multipole analysis using group theory}
If there is an acoustic resonator with a defined symmetry, then the equation describing the system is invariant under the symmetry transformations of the group. For the compact resonator which can be considered to be a perturbation of a spherical resonator~\cite{Bogdanov_Koshelev_Kapitanova_Rybin_Gladyshev_Sadrieva_Samusev_Kivshar_Limonov_2019,PhysRevA.90.013834} its eigenmodes are similar to the sphere's eigenmodes. 

Let us consider a particular eigenmode of an acoustic resonator which is described by a complex amplitude of pressure $p(\vb{r})$. We note that real observed field are defined as $p(\vb{r}, t) = \Re [p(\vb{r})e^{- i \omega t}]$. 
We can write a multipole decomposition as a sum of scalar spherical functions~\cite{williams1999fourier}:
\begin{equation}
    p(\vb{r}) = \sum_{p,\ell,m} c_{p\ell m}(r) Y_{p\ell m}(\vartheta, \varphi),
    \label{phi_sum}
\end{equation}
where the summation is taken over all the indexes  as follows  $\sum_{p,\ell,m} \equiv \sum_{p=e,o} \sum_{\ell=0}^\infty \sum_{m=-\ell}^\ell$, $r = |\vb{r}|$ is the magnitude of the radius vector, $\vartheta$ is the polar angle and $\varphi$ is the azimuthal angle in the spherical coordinate system (see Fig~\ref{SF_table} inset). Indeed, the real spherical functions form a basis in space with scalar product given as integral of the solid angle
\begin{equation}
    \langle Y_{p\ell m}, Y_{p'\ell'm'} \rangle = \int_{4\pi}  Y_{p\ell m} Y_{p'\ell'm'} d\Omega = \delta_{p\ell m}^{p'\ell'm'}
    \label{scalar_product}
\end{equation}
where $\delta_{p\ell m}^{p'\ell'm'}$ is a Kronecker delta, 
$\int_{4\pi} d\Omega \equiv \int_{\varphi=0}^{2\pi} \int_{\theta=0}^{\pi} \sin{\theta} d\theta d\varphi$. The multipole content of the mode $c_{plm}$ can fully describe the mode properties, however we leave the discussion of multipole series convergence and accuracy out of the scope of this paper~\cite{Augui__2016} saying that the precision is high enough for our purposes.

According to the Wigner theorem a specific eigenmode $p(\vb{r})$ is transformed under particular irreducible representation of the resonator's symmetry group. Thus, in the expansion~\eqref{phi_sum}, only spherical functions which are transformed under this irreducible representation are presented. There cannot be any spherical functions transformed under different representation in the expansion. Accordingly, the problem of multipole expansion of eigenmodes of the resonator is reduced to determining from symmetry considerations under which irreducible representation a particular mode is transformed and finding a set of spherical harmonics $ Y_{p\ell m}$, transformed through each other under the same one. To find such a set, the projection operator on the irreducible representation is introduced $\hat{P}_\alpha$~\cite{7031881, Knox_Gold_1964}:
\begin{equation}
    \hat{P}_\alpha = \frac{n_\alpha}{|G|}\sum_g \chi^*_\alpha(g)\hat{D}(g)
\end{equation}
where 
$|G|$ is the order of the group,
$g$ is the group element,
$\alpha$ is the number of irrep,
$n_\alpha$ is the dimension of irrep,
$\chi^*_\alpha(g)$ is the character of $g$,
and $\hat{D}(g)$ is the transformation operator.

\begin{figure}[t]
            \includegraphics[width = 0.47\textwidth]{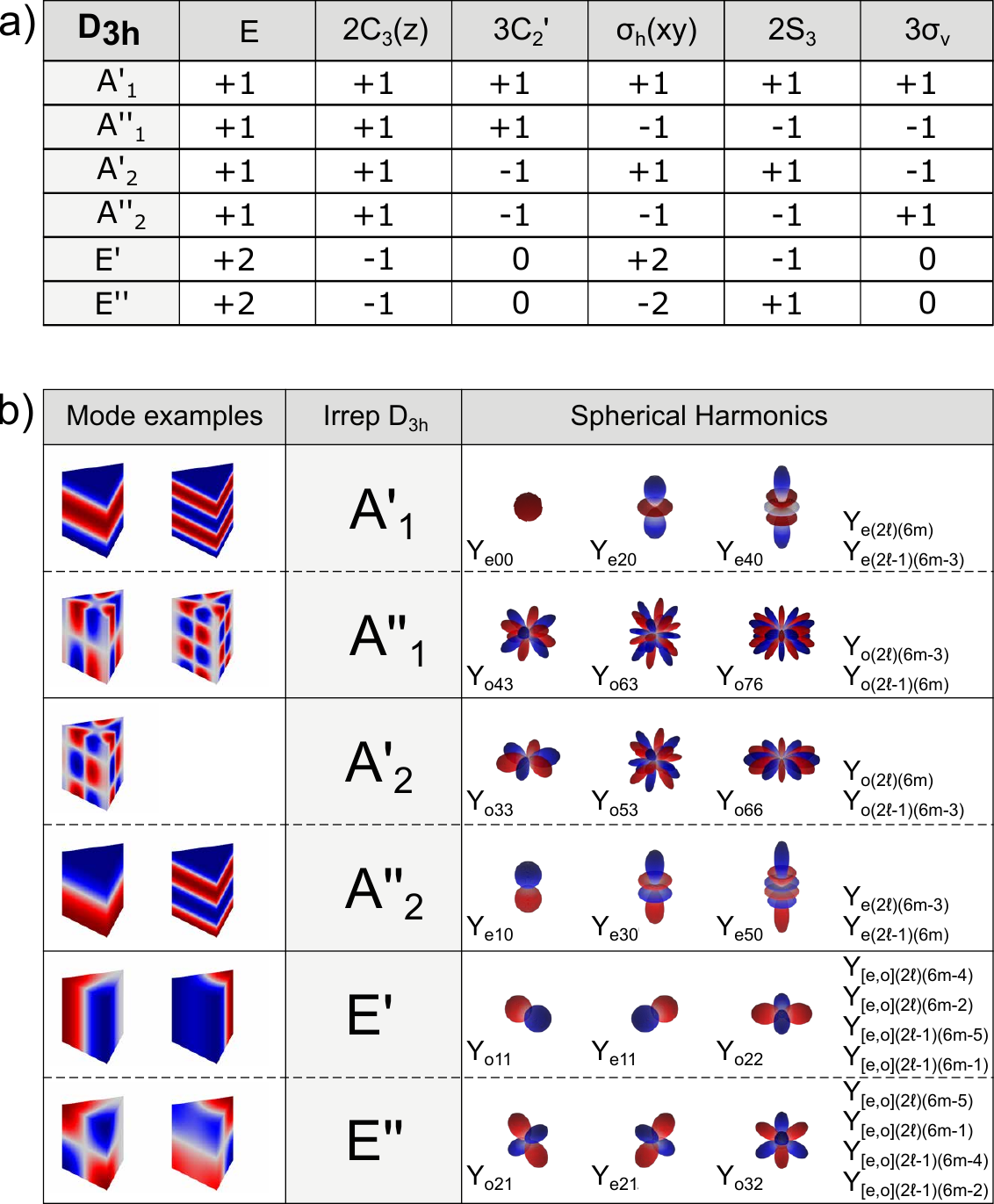} 
\caption{a) Character table for symmetry group $D_{3h}$ group. b) Table of multipole composition of eigenmodes for a closed acoustic resonator of symmetry group $D_{3h}$: examples of modes transformed under particular irreducible representation (first column) and their multipolar content (third column).}\label{d3htabl}
\end{figure}

The  operator $\hat{P}_\alpha$ projects arbitrary basis functions onto a linear combination of functions which are transformed under particular irreducible representation $\alpha$~\cite{olver_1999}.
Commonly, this procedure allows for identification of the basis functions. However,  for the case of spherical functions set we can simply use the already obtained results and address to the readymade character tables~\cite{paper,spice}. With this, knowing  that a particular eigenmode transforms under irrep $\alpha$, one can immediately  determine the multipole  composition of this mode  by simply finding the spherical functions, which are also transformed under the same irrep $\alpha$. This set will determine the non-zero components in the expansion Eq.~(\ref{phi_sum}). For the resonators of  $D_{3h}$ symmetry, the multipolar content is shown in Fig.~\ref{d3htabl}. The first column shows the results of numerical simulation  obtained using an eigenmode solver in COMSOL Multiphysics\texttrademark, where the color denotes the  acoustical pressure at the surface of the resonator.

\begin{figure}[ht]
\begin{tabular}{p{0.49\textwidth}}
     \includegraphics[width = 0.48\textwidth]{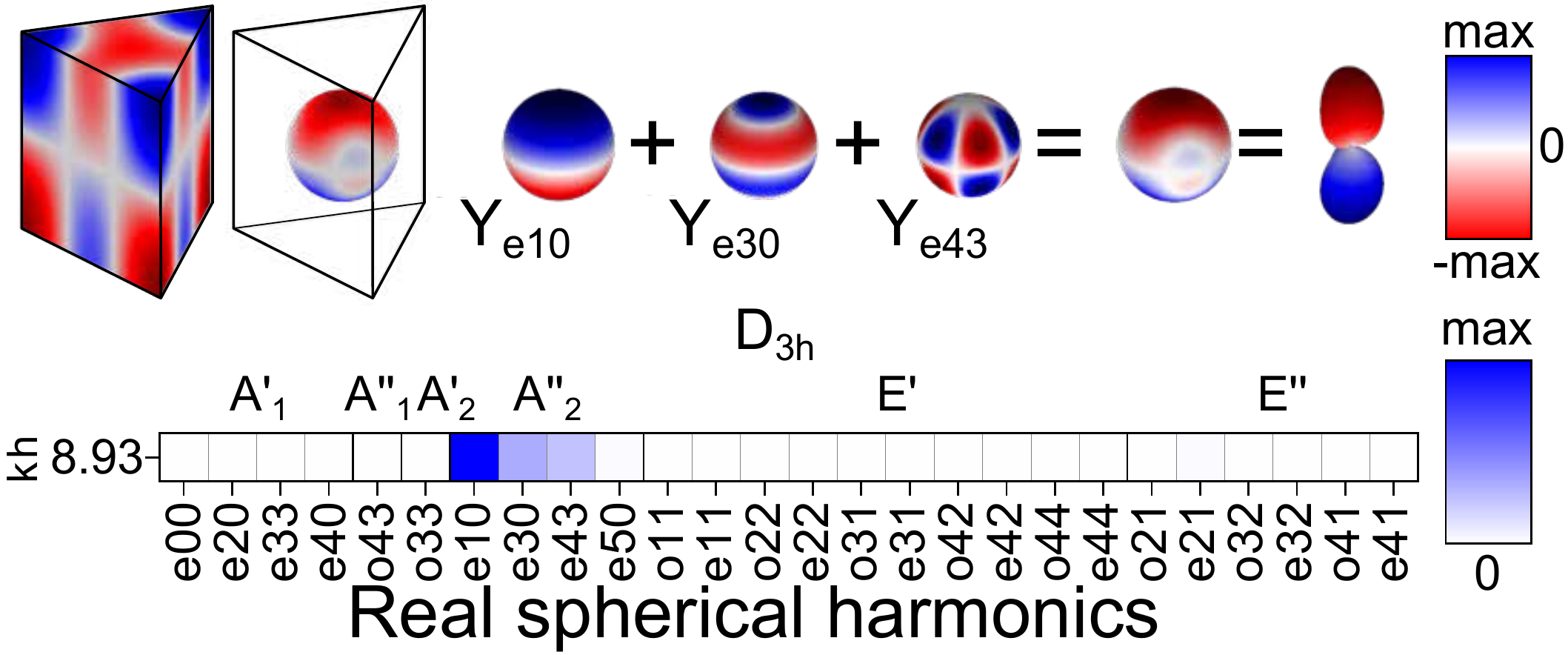}\\
\end{tabular}
 
  \caption{Illustration of the analysis of the multipole composition of a particular eigenmode of a closed resonator. The integration of the pressure function of the eigenmode  multiplied by the spherical function was performed over  an auxiliary  sphere inside the structure. The nonzero value of the integral (filled  colored cells in the  table) indicates that such spherical function is included in the eigenmode expansion with a certain coefficient.}
    \label{Example}
\end{figure}

\subsection{Numerical simulations}
The group theory approach provides  the non-zero multipole coefficients in the expansion Eq.~(\ref{phi_sum}). The quantitative analysis of these coefficients is usually carried out  by direct multipole decomposition of the modes of open resonators\cite{gladyshev}. Here, for a closed resonator under consideration, we are rather interested in testing the predicted non-zero multipole components basing on the numerical simulations.   For that, firstly we numerically compute the pressure distributions of the eigenmode $\psi(\vb r)$ inside the resonator with help of commercially available  COMSOL Multiphysics\texttrademark\quad software. Then, the eigenmode function is multiplied by spherical harmonics $Y_{p\ell m}$ and integrated over a sphere surface embedded inside the resonator and with center matching the resonator center of symmetry as shown in  Fig.~\ref{Example}:
\begin{multline}
    \langle \psi, Y_{p\ell m} \rangle = \sum_{p',\ell',m'} c_{p'\ell'm'} \int_{4\pi} Y_{p'\ell'm'} Y_{p\ell m} d\Omega = \\ = \sum_{p',\ell',m'} c_{p'\ell'm'} \delta_{p\ell m}^{p'\ell'm'} = c_{p\ell m}
    \label{multipole_exp_c}
\end{multline}


A non-zero result of the integration represents the fact that the spherical function is included to the eigenmode expansion (see  Fig.~\ref{Example} for a particular eigenmode of a prism resonator). The result will depend on the size of the sphere however we are interested in zero values of the coefficients due to symmetry restriction which will stay zero for any sphere size. 

\begin{figure}[h!t]
    
    \includegraphics[width=0.48\textwidth]{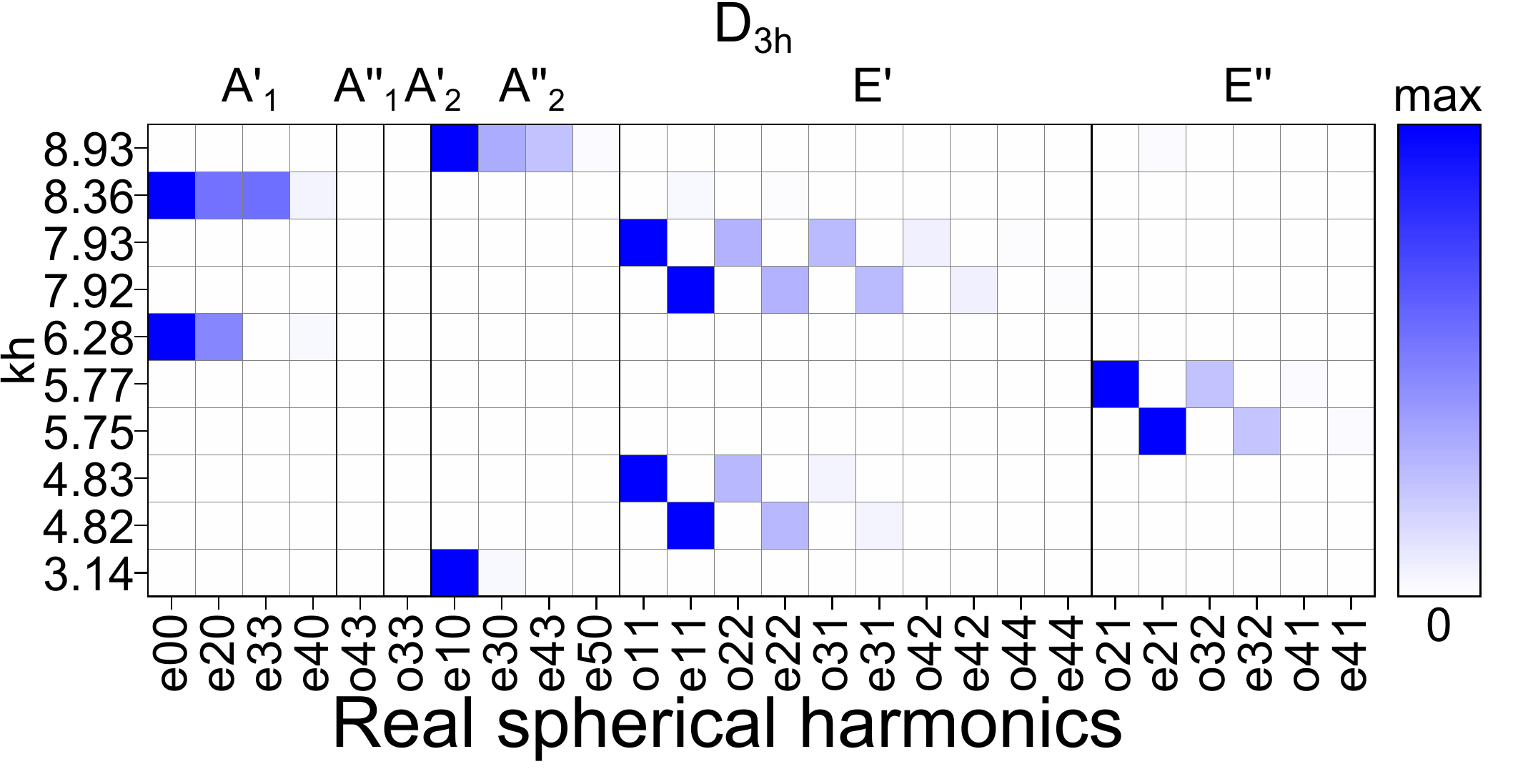}
    \caption{A part of the infinite table of coefficients at spherical functions for the symmetry group $D_{3h}$. On the vertical axis, the product of the wavevector $k$ of eigenmodes by the height $h$ of the prism is given. The color saturation corresponds to the value of the coefficient $c_{p\ell m}$ in the expansion. The checkerboard arrangements correspond to degenerate modes transformed under two-dimensional representations of $E'$, $E''$. The length of the horizontal base of the prism (object of symmetry group $D_{3h}$) is 44.68 mm, and its height is 50 mm. However geometrical properties are unimportant to this end.} 
    \label{table_example}
\end{figure}

\begin{figure}[h!]
    \centering
    \includegraphics[width=0.35\textwidth]{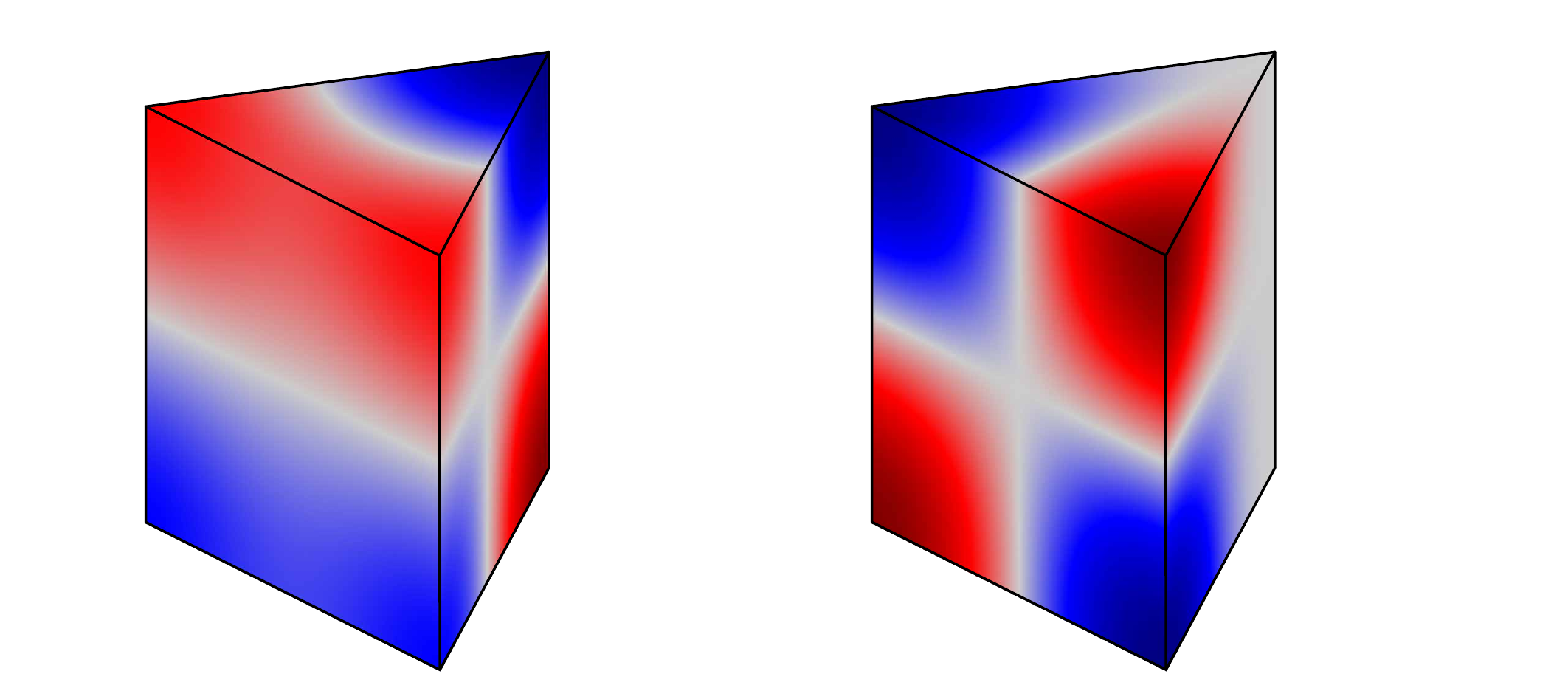}
     \caption{Degenerated modes transformed under the irreducible representation $E''$.}
    \label{deg_mods}
\end{figure}

Fig.~\ref{table_example} summarizes the results of the symmetry analysis for different eigen modes of $D_{3h}$ group symmetry. It shows that for each eigenfrequency coefficients deviate from zero in one representation only. Moreover, there are degenerated modes in two-dimentional representations: two lines correspond to a single frequency, and coefficients are in blocks or chequerwise. 
While finding the degenerate modes, the numerical solver oftenly  selects them in arbitrary manner, however by slightly violating the symmetry, we can force it to select a particular linear combination. The case of the checkboard positioning corresponds to the case, where in each line every spherical function is either odd or even (Fig.~\ref{deg_mods}), therefore one of the two degenerate modes is even when reflected in $y=0$ plane, and the second one is odd. The chequerwise distribution   can be achieved artificially by stretching the shape out along the $x$-axis. In this case modes would not be truly degenerated due to the lower symmetry of the system, but their frequencies would be close enough.

\begin{figure*}[ht]
    \includegraphics[width = 1\textwidth]{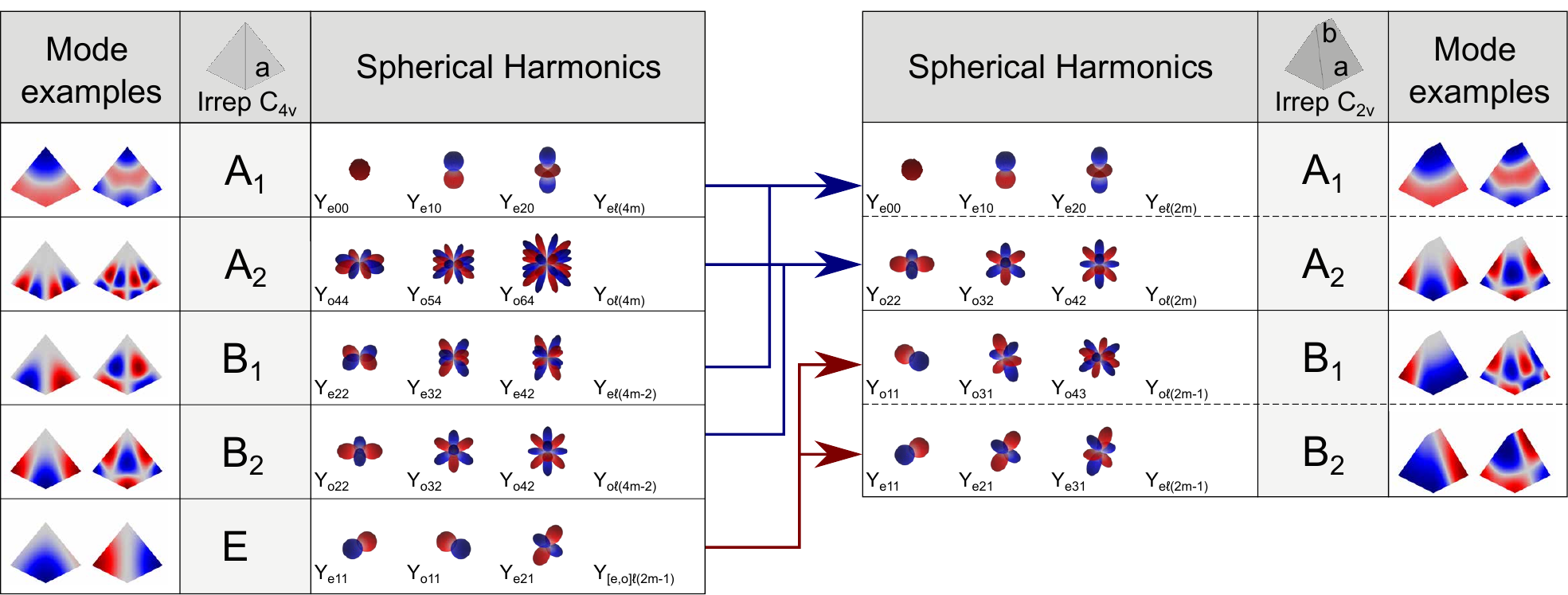}\\
    \caption{Two symmetry groups $C_{4v}$, $C_{2v}$, one of which is a subgroup of the other. Each irreducible representation corresponds to explicit examples of eigenmodes of the resonator transformed under given irreducible representation and expressed by a linear combination of spherical functions transformed under the same one. The blue arrows illustrate the reduction of the number of irreducible representations with decreasing symmetry, and the red arrows show the cancellation of degeneracy with breaking the rotational symmetry around the $z$-axis by 90 degrees. 
    Geometry of studies closed resonators: $a=50$ mm, $b=20$ mm, overall height is 50 mm. For each resonator: density $\rho = 1190 \; \text{kg}/\text{m}^3$, speed of sound $c = 2500\; \text{m}/\text{c}$
    However geometrical properties are unimportant to this end.}
    \label{C4vC2v}
\end{figure*}
\subsection{Eigenmodes decomposition of closed resonators of different symmetry groups} \label{sec:Eigenmodes_decomposition}
As a next step, we have performed the classification of resonators of other  symmetry groups of $D_{2h}$,  $D_{3h}$, $D_{3d}$, $D_{4h}$, $D_{6h}$, $D_{\infty h}$, $C_{2v}$, $C_{3v}$, $C_{4v}$, $C_{6v}$, $C_{\infty v}$ with customized geometrical parameters (see Appendix \ref{sec:appendix0}). The tables for all symmetries are given in Appendix \ref{sec:appendix1} in the form similar to  Fig.~\ref{d3htabl}. 


\subsection{
Multipole content in the resonators of decreased symmetry}

In this part, we would like to focus attention on an interesting behavior when lowering the symmetry of a resonator and correspondent multipoles content evolution. In Fig.~\ref{C4vC2v} the multipoles set of pyramid resonator ($C_{4v}$-group symmetry)  is shown. Under particular transformation one can decrease its symmetry to $C_{2v}$, which is a subgroup of $C_{4v}$.  When the symmetry is being lowered the number of irreducible representations is reduced: two one-dimensional representations are merged, forming a different one-dimensional  representation ~\cite{dresselhaus_dresselhaus_jorio_2008}. Moreover, degeneracies are cancelled when the rotational symmetry around the $z$-axis is broken
: two-dimensional representation turns to two one-dimensional, in other words two degenerated modes turn into two modes with different energies. In the following section, we will demonstrate the effect it has on the spectral features of acoustic wave scattering.

\section{Acoustic wave scattering} 
\label{sec:scattering}
\begin{figure*}[ht]
    \centering

    \includegraphics[width = \textwidth]{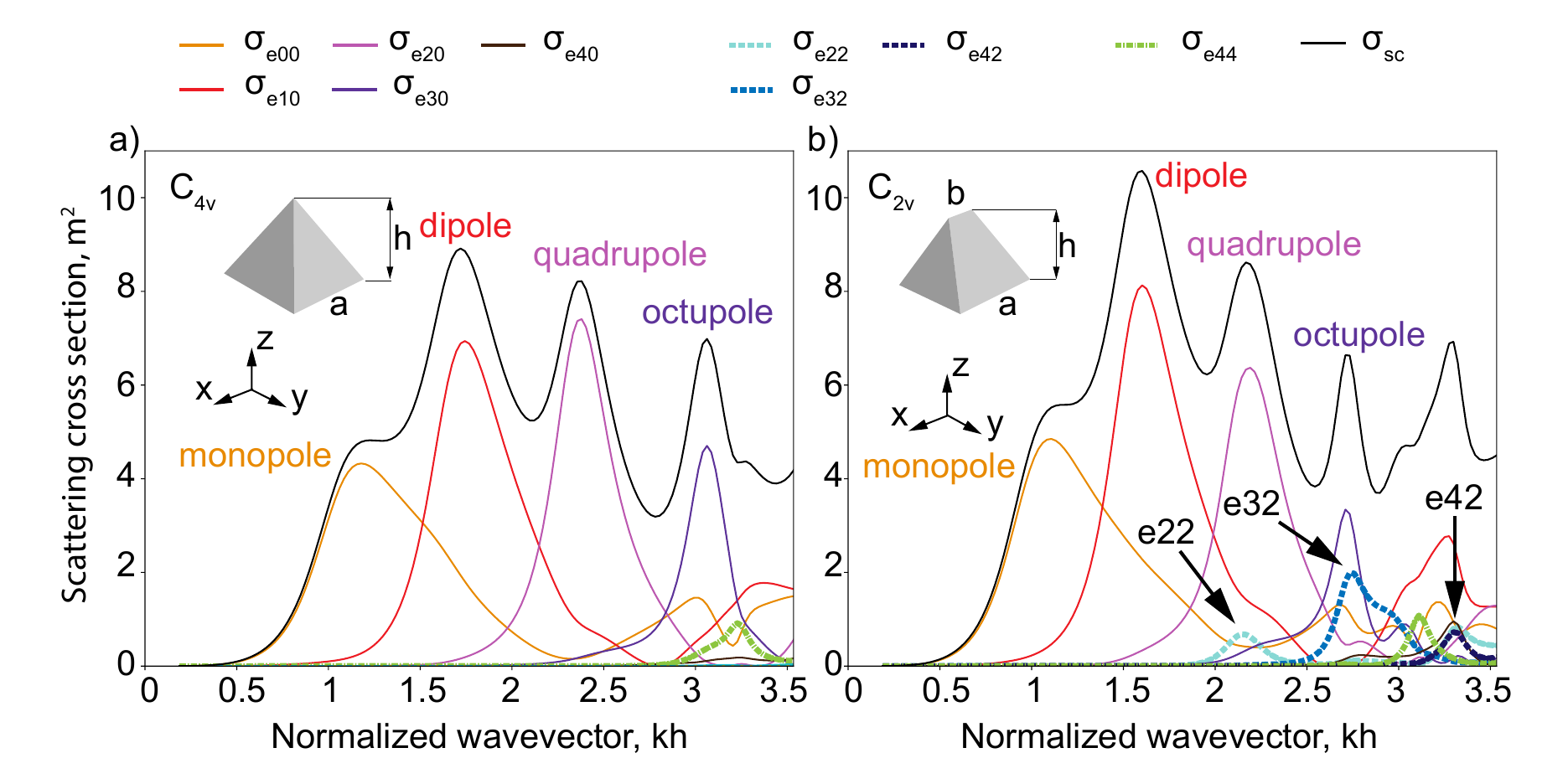}
    \caption{ Scattering cross section expansion of two objects with $C_{2v}$ (a) and $C_{4v}$ (b) symmetries. Plane wave propagates along $z$-axis and therefore contains spherical harmonics with $m=0$ only in its expansion. 
    (b) The figure illustrates how slight change of a symmetry alters the scattering spectrum multipolar content. Additional contributions of $Y_{e\ell2}$ are shown. 
    Geometry parameters are $a=1.5\;$m, $b=0.6\;$m, overall height $h=1.5\;$m; parameters of the resonators: speed of sound $c=2\;$m/s, density $\rho=1/\sqrt{6}\;$kg/m$^3$; parameters of the host medium: speed of sound $c_0=1\;$m/s, density $\rho_0=1\;$kg/m$^3$} 
    \label{scattering}
\end{figure*}

The scattering of an acoustic wave on the  resonator having particular set of eigenmodes results in appearance of resonant features in the scattering spectrum. At this stage, the information on the particular  multipole content  of resonators can be extremely helpful to analyze and predict the spectral response of the resonant object. 

\subsection{Multipoles expansion}
Radiation pressure of a  plane acoustic wave, propagating along $z$-axis is  in the form of~\cite{williams1999fourier,PhysRevLett.123.183901,doi:10.1063/1.5058149,doi:10.1121/1.1906621, MARTIN201963}
\begin{equation}
p^{i}=p_{0} e^{i k r \cos \theta}=\sum_{\ell=0}^{\infty} p_{\ell} j_{\ell}(k r) Y_{e\ell 0}(\cos \theta)
\end{equation}
where $p_{\ell}=p_0i^\ell \sqrt{4\pi (2\ell+1)}$, $p_0$ is the incident wave amplitude, $j_\ell (kr)$ is the  spherical Bessel function. By virtue of symmetry, there are solely spherical harmonics $m=0$ presented in the expansion, however for the scattered wave there might be all components in the expansion:

\begin{equation}
p^{s}(\mathbf{r}, \omega)=\sum_{p,\ell, m} a_{p\ell m }(\omega) h_{\ell}^{(1)}(k r) Y_{p\ell m}(\cos \theta),
\end{equation}
where $h_{\ell}^{(1)}$ is the Spherical Hankel function of the first kind~\cite{PhysRevE.99.063004}.
The scattering cross section can be expressed through the scattering coefficient as:
\begin{equation}
\sigma_{sc}=\frac{1}{k^2}\sum_{p,\ell,m} |a_{p\ell m }(\omega)| ^2 = \sum_{p,\ell,m} \sigma_{p\ell m }(\omega)
\label{sigma_sc}
\end{equation}
where $ \sigma_{p\ell m }(\omega) = \frac{1}{k^2}|a_{p\ell m }(\omega)| ^2$.
The scattering cross section also can be obtained by direct integration of the scattered wave energy flux over any surface surrounding the resonant object, for instance a sphere of radius $R$,  
\begin{equation}
    \sigma_{sc} = \frac{R^2}{I_0}\int_{4\pi} \frac{1}{2}  \Re({p^s}^* v_r^s)d\Omega
    \label{sigma_sc2}
\end{equation}
where $I_0$ is the incident wave intensity, $v_r^s$ is the component of the local velocity of the scattered wave~\cite{Gong_Li_Chai_Zhao_Mitri_2017}. The answer does not depend  on the integration sphere radius since the  energy flux  of  the scatterer power $\frac{1}{2} \Re p^{s*} \vb{v}^s \propto 1/R^2$.
The multipole expansion coefficients can be found by projecting the pressure of the scattered wave onto the corresponding spherical harmonic: 
\begin{widetext}
\begin{equation}
    \langle p^{s}(\mathbf{r}, \omega), Y_{p\ell m} \rangle =\sum_{p',\ell',m'} a_{p'\ell'm'}(\omega)  h_{\ell'}^{(1)}(k r) \int_{4\pi} Y_{p'\ell'm'} Y_{p\ell m} d\Omega  = \sum_{p',\ell',m'} a_{p'\ell'm'}(\omega) h_{\ell'}^{(1)}(k r) \delta_{p\ell m}^{p'\ell'm'} =a_{p\ell m}(\omega)  h_{\ell}^{(1)}(k r)
\label{a_coef}
\end{equation}
\end{widetext}

If  the spherical function $Y_{p\ell m}$ is included both in the incident wave expansion and in an eigenmode's expansion, than this  eigenmode will be  excited along with all others, which are transformed by the same irrep as  $Y_{p\ell m}$. Therefore in the scattered wave the whole basis of spherical harmonics transformed under that irreducible representation is constituted.  The incident plane wave propagating along the $z$-axis  contains only spherical functions with $m=0$ and, thus, it excites  all modes which contain at least one spherical function with $m=0$.

\subsection{Acoustic scattering results}

Let us now consider a particular case of wave scattering over an open acoustic resonator of  $C_{2v}$ and $C_{4v}$ symmetry groups.  In accordance with  the multipole analysis results mentioned earlier (Fig.~\ref{C4vC2v}), the incident wave excites  only one irreducible representation, $A_1$. However,  irreducible representation $A_1$  for $C_{2v}$ symmetry is a result of merging irreducible representations $A_1$ and $B_1$ of $C_{4v}$ symmetry group, therefore irreducible representation $A_1$ for $C_{2v}$ symmetry group corresponds to $Y_{e\ell (2m)}$ spherical functions, while in $C_{4v}$ group there are only $Y_{e\ell (4m)}$ functions present. As a result, the multipole expansion of the wave scattered by the object with $C_{2v}$ symmetry shows contributions of modes missing in the scattering on an object with $C_{4v}$ symmetry  (Fig.~\ref{scattering}). The used parameters of models and environment are presented in Appendix \ref{sec:appendix1}.

 \subsection{Analogy to optics}
 
 Recently, the group symmetry analysis of eigen modes and their multipolar content of subwavelength optical resonator has been carried out in optics\cite{gladyshev}, where multipolar theory is a powerful tool for describing optical response of resonant nanostructures, predict, and design their optical properties\cite{Raab2005, Smirnova_Kivshar_2016, Sadrieva2019, Liu2020}.  The main difference between the optics and acoustics is that the electromagnetic fields are described by vector functions which adds a polarization degree of freedom. When considering the farfield multipolar content of the modes, one should  consider two types of vector spherical harmonics --- electric and magnetic ones. However, despite of complexity of electromagnetic modes, one can note that content of the acoustic and optical modes is rather similar \cite{gladyshev} as the electric multipoles behave almost exactly as scalar spherical harmonics. The magnetic harmonics have  opposite behavior under inversion and reflection, and can excite, for example, modes which transform under irreducible representations $A_{2g}$ and $A_{2u}$ in cylinder, which is impossible in acoustics. Another important difference between acoustic and optical scattering is that the plane wave, incident along $z$-axis, has only $m=1$ due to vector nature of the electromagnetic fields~\cite{bohren2008absorption} and, thus,   transforms under a different irreducible representation. Cylindricaly symmetric optical fields with $m=0$ can be achieved in vector beams. We summarize the correspondence between acoustics and optics in Table~\ref{tabl}.

 \begin{table}[h!]
\begin{tabular}{|l|c|lll}
\cline{1-2}
\multicolumn{1}{|c|}{Acoustics} & Optics                                         &  &  &  \\ \cline{1-2}
Scalar spherical functions $Y$      & Electric multipoles $\vec N$                            &  &  &  \\ \cline{1-2}
\multicolumn{1}{|c|}{-}         & Magnetic multipoles $\vec M$                            &  &  &  \\ \cline{1-2}
m=0 for a plane wave k||z       & \multicolumn{1}{l|}{m=1 for a plane wave k||z} &  &  &  \\ \cline{1-2}
\end{tabular}
\caption{Relation between the multipole analysis  in  acoustics and optics.}\label{tabl}
\end{table}
\section{CONCLUSION} 
\label{sec:conclusion}

Finally, in conclusion, we applied the machinery of group theory in order to classify and analyze the modes of subwavelength acoustic resonators in a way similar to nanophotonics. We considered the resonators eigenmodes of several symmetry groups ($D_{2h}$, $D_{3h}$, $D_{3d}$, $D_{4h}$, $D_{6h}$, $D_{\infty h}$, $C_{2v}$, $C_{3v}$, $C_{4v}$, $C_{6v}$, $C_{\infty v}$) and presented the classification tables of their eigenmodes and the multipolar content for each class. The proposed approach of multipole classification approach can be extended  to any resonator  shape and material. We have connected the multipolar classification with the acoustic scattering problem, since  knowing only the multipolar structure of the incident wave and the symmetry group of the resonator one can  predict the exact multipolar composition of the scattered field. We studied in detail the lift of degeneracy in a  resonator of $C_{4v} \to C_{2v}$ symmetry and traced the evolution of multipolar content as well as reconfiguration the scattering spectrum. In particular,  we showed that  symmetry decreasing leads to expanding the range of  multipoles in the scattered field. We believe, that our result will find an application in rapidly developing acoustic of metamaterials and  metaatoms.

\begin{acknowledgments}
	The authors would like to thank Yuri Kivshar for the fruitful discussions. This work was supported by Russian Science Foundation (project 20-72-10141). KF acknowledges support from the Foundation for the Advancement of Theoretical Physics and Mathematics “BASIS” (Russia)
\end{acknowledgments} 
\FloatBarrier
\bibliography{literatur}

\appendix
\onecolumngrid
\newpage
\section{COMSOL Multiphysics\texttrademark\texorpdfstring{\,}{Lg} model}\label{sec:appendix0}
We used Eigenvalue Solver to observe electromagnetic fields of closed resonators. We established spherical domain inside the resonator (Fig. \ref{Example}) to estimate $c_{p\ell m}$ coefficient in the multipole expansion (\ref{multipole_exp_c}). We integrate pressure and spherical harmonics (\ref{scalar_product}) over the surface of the sphere:
\begin{equation}
    c_{p\ell m} = \int_{4\pi} p(\mathbf{r})  Y_{p\ell m} d\Omega
\end{equation}
We used a built-in complex form of spherical harmonics to get a real solution (\ref{sh_comlex_to_real}).

For open resonator we used Frequency Domain Solver. The model geometry is a sphere with a Perfect Matched Layer (PML) domain and resonator at the center. To get $a_{p\ell m}(\omega)$ (\ref{a_coef}) we integrated scalar product of far-field scattering pressure $p^{s}$ and spherical harmonics over concentric parametric surface with a radius $R$ outside of the resonator, but inside the PML sphere:
\begin{equation}
    a_{p\ell m}(\omega) = \frac{1}{h_{\ell}^{(1)}(k R)}\int_{4\pi} p^{s}(\mathbf{r}, \omega)  Y_{p\ell m} d\Omega
    \label{eq:alm_comsol}
\end{equation}
We also note that in Eq.~\eqref{eq:alm_comsol} one should use $h_\ell^{(2)}$ in COMSOL Multiphysics\texttrademark\ instead of $h_\ell^{(1)}$. This comes from the fact the outgoing wave should have the asymptotics as $e^{-ikr}$ once the $e^{+i \omega t}$ convention is used, which is the case for the COMSOL  Multiphysics\texttrademark. Sum of squared absolute values of coefficients equals $\sigma_{sc}$ (\ref{sigma_sc}).

We used the same concentric parametric surfaces to get $\sigma_{sc}$ by (\ref{sigma_sc2}). The following expression was used as an integrand in COMSOL Multiphysics\texttrademark\  model (built-in acoustics interfaces notation)
\begin{verbatim}
0.5*realdot(acpr.p_s,(
	(-(d(acpr.p_s,x))*x)+
	(-(d(acpr.p_s,y))*y)+
	(-(d(acpr.p_s,z))*z)
)/(acpr.rho_c*acpr.iomega))
\end{verbatim}

\setcounter{figure}{0}
\section{Models and geometry of studies closed resonators}\label{sec:appendix1}
\begin{figure}[H]
    
    \centering
        \begin{tabular}{c}
            \includegraphics[width = 0.4\textwidth]{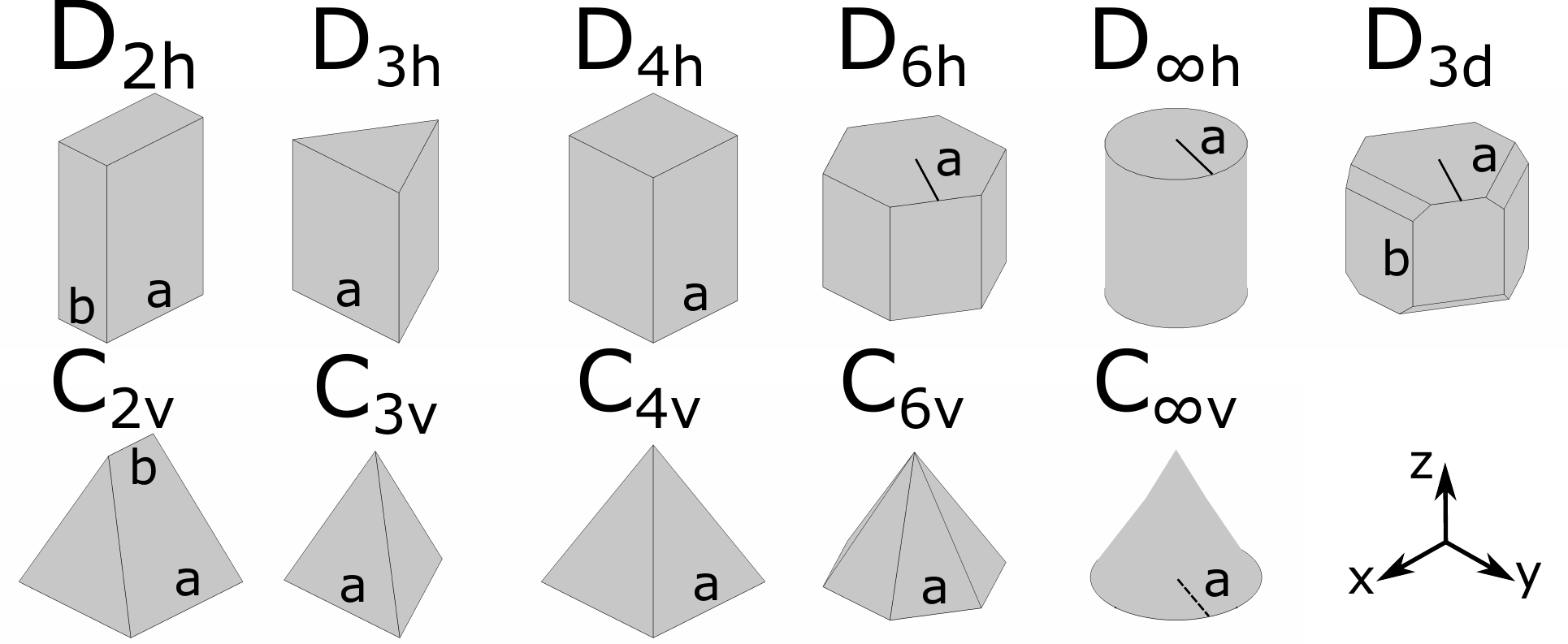}
        \end{tabular}
        
        \begin{tabular}{c|cccccc}
            \hline
             Parameter &$D_{2h}$ & $D_{3h}$ & $D_{4h}$ & $D_{6h}$ & $D_{\infty h}$ &  $D_{3d}$ \\\hline
             $h$, mm & 60 & 50 & 40 & 40 & 30 & 40 \\
             $a$, mm & 40 & 44.68 & 25 & 25 & 12.5 &  25 \\
            $b$, mm & 20 & &  & & & 30 \\\hline \hline
    
            Parameter &$C_{2v}$ &$C_{3v}$ & $C_{4v}$ & $C_{6v}$ & $C_{\infty v}$ & \\\hline
            $h$, mm & 50 & 50 & 50 & 37.5 & 21.65 &\\
            $a$, mm & 50 & 50 & 50 & 50 & 12.5 &\\
            $b$, mm & 20 &   &    &     &     &\\\hline
        \end{tabular}

        \label{fig:sizes}
        \caption{Geometry of studies closed resonators. $h$ -- overall height. For each resonator: density $\rho = 1190 \; \text{kg}/\text{m}^3$, speed of sound $c = 2500\; \text{m}/\text{c}$}
\end{figure}
\begin{figure}[H]
\centering
    \begin{tabular}{c|ccc}
            \hline
             Parameter &$C_{4v}$ & $C_{2v}$ & Host \\\hline
             $h$, m & 1.5 & 1.5 & \\
             $a$, m & 1.5 & 1.5 & \\ 
             $b$, m &  & 0.6  &\\
             $c$, m/s & 2 &  2 & 1\\
             $\rho$, kg/m$^3$ & $1/\sqrt{6}$ & $1/\sqrt{6}$ & 1\\ \hline
\end{tabular}

    \caption{Parameters of the studied resonators and media in the Mie scattering problem}
\end{figure}

\section{Eigenmodes' multipole decomposition results}\label{sec:appendix2}
\begin{figure*}[h!]
\caption{Tables of irreducible representations and examples of eigenmodes transformed under them, and tables of characters of irreducible representations for symmetry groups  $D_{4h}$, $D_{2h}$, $D_{3d}$, $D_{3h}$ }
\label{d4h}
\begin{tabular}{p{0.45\textwidth}p{0.45\textwidth}}
        \vspace{0pt}
        \begin{tabular}[t]{c}
            \includegraphics[width = 0.45\textwidth]{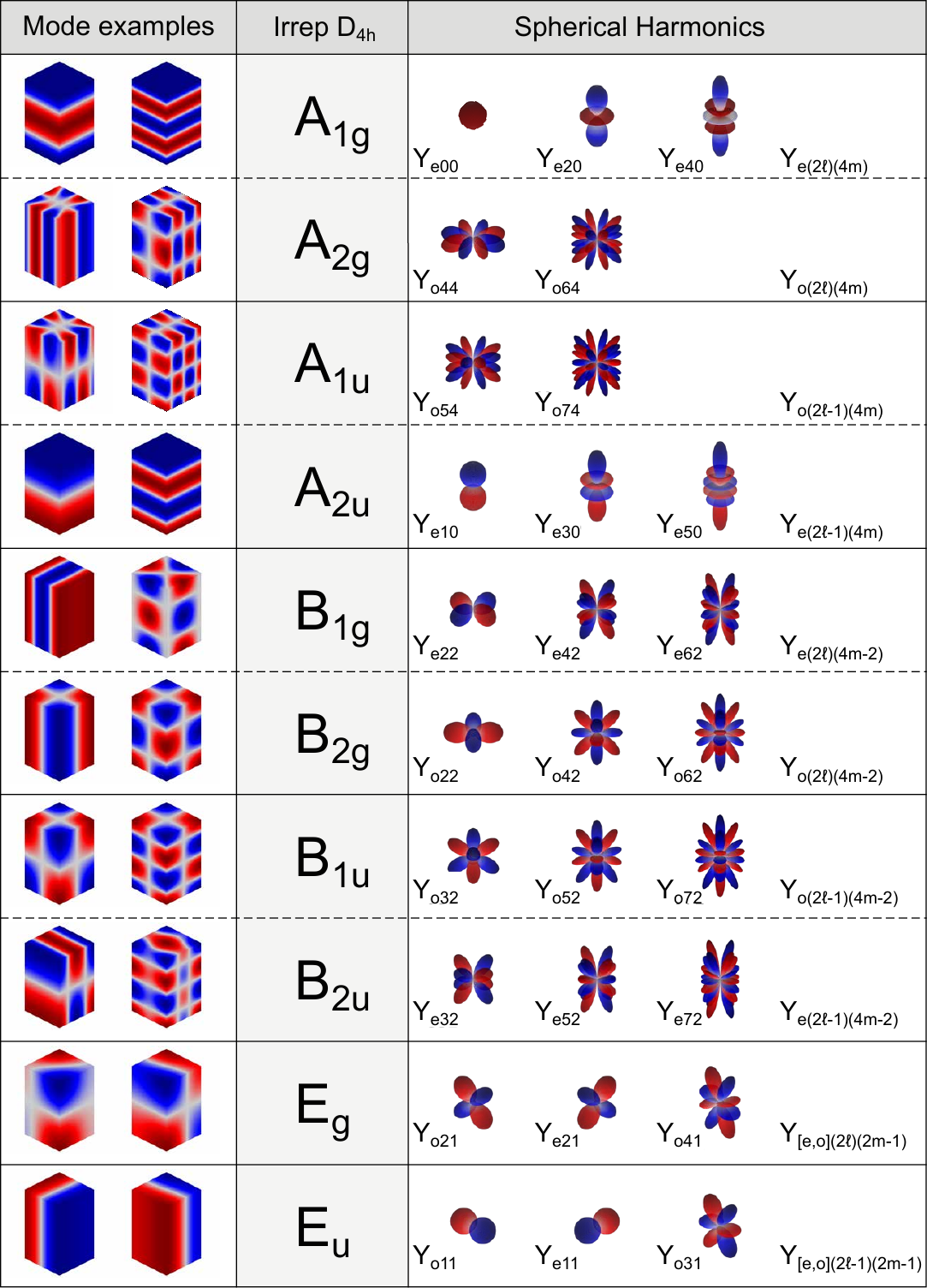}
        \end{tabular}
        &
        \vspace{0pt}
        \begin{tabular}[t]{r}
            \includegraphics[width = 0.45\textwidth]{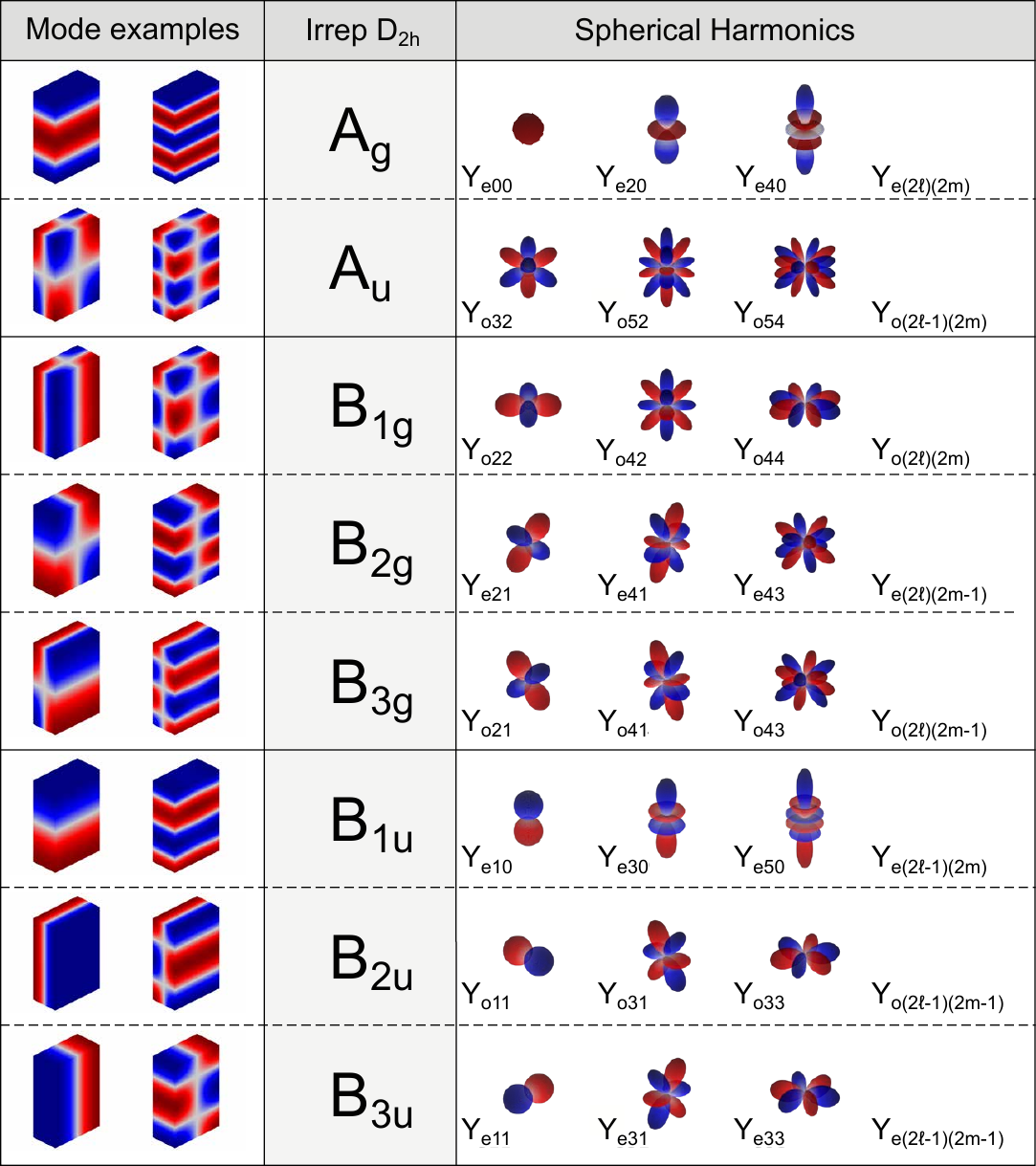}
            \end{tabular}
\\
        \vspace{0pt}
        \begin{tabular}[t]{c}
            \includegraphics[width = 0.45\textwidth]{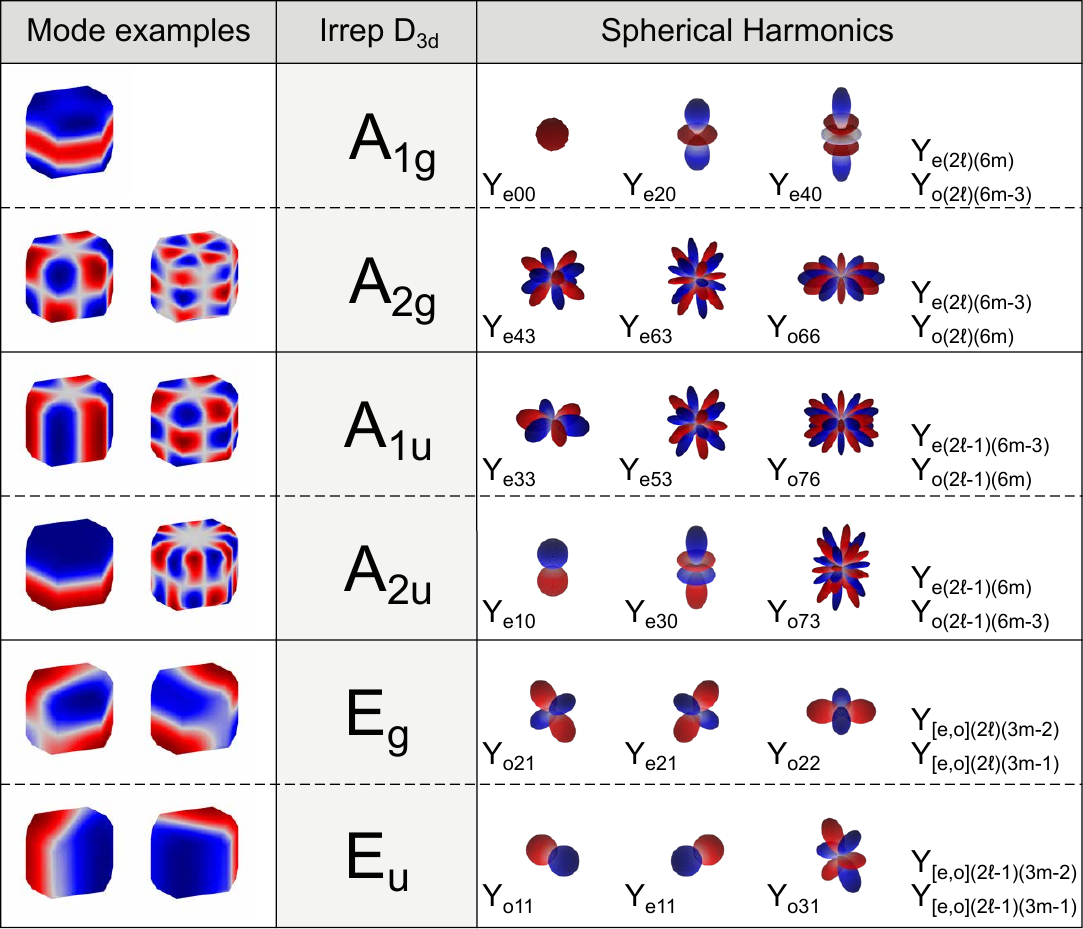}
        \end{tabular}  
        &
        \vspace{0pt}
        \begin{tabular}[t]{r}
            \includegraphics[width = 0.45\textwidth]{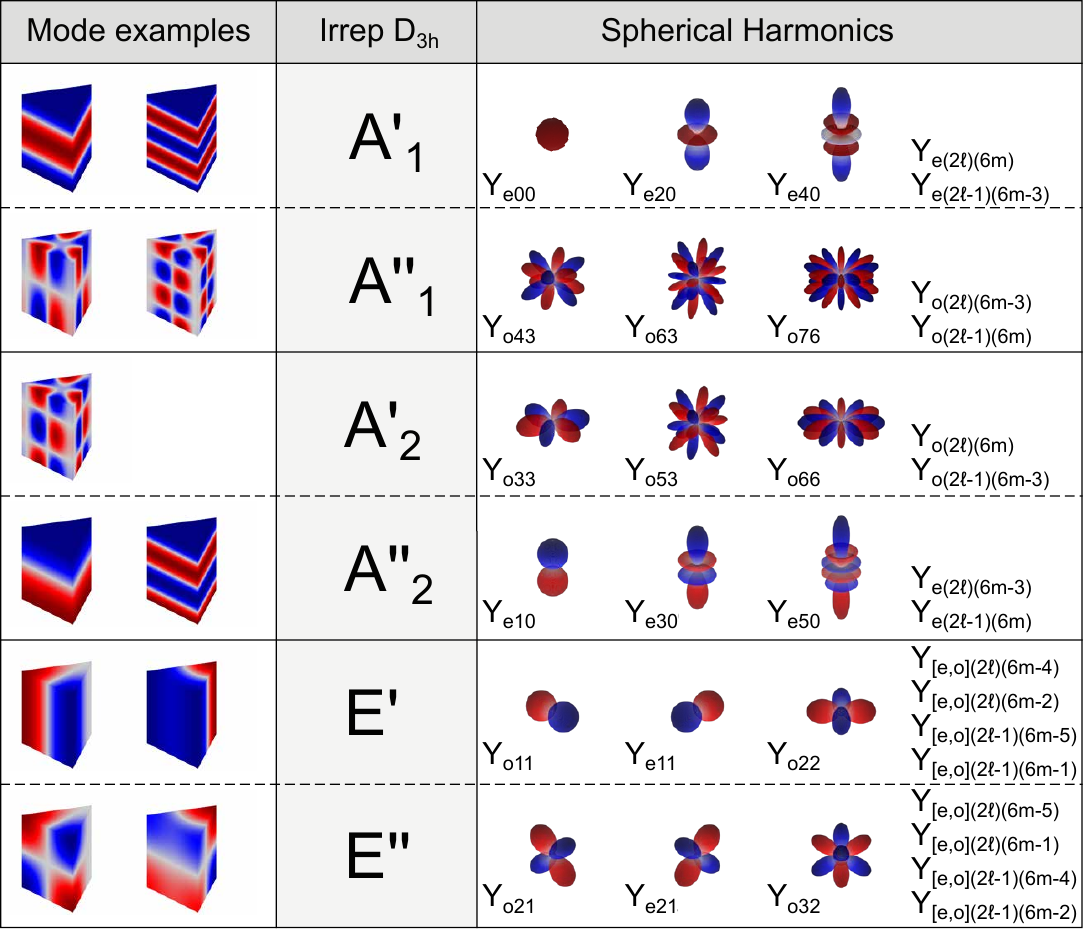}
        \end{tabular} 
\end{tabular}
\end{figure*}

\begin{figure*}[hp]
\caption{Tables of irreducible representations and examples of eigenmodes transformed under them, and tables of characters of irreducible representations for symmetry groups  $C_{\infty v}$, $C_{6v}$, $D_{\infty h}$, $D_{6h}$ }
\label{d4h}
\begin{tabular}{p{0.5\textwidth}p{0.49\textwidth}}
        \vspace{0pt}
        \begin{tabular}[t]{c}
            \includegraphics[width = 0.49\textwidth]{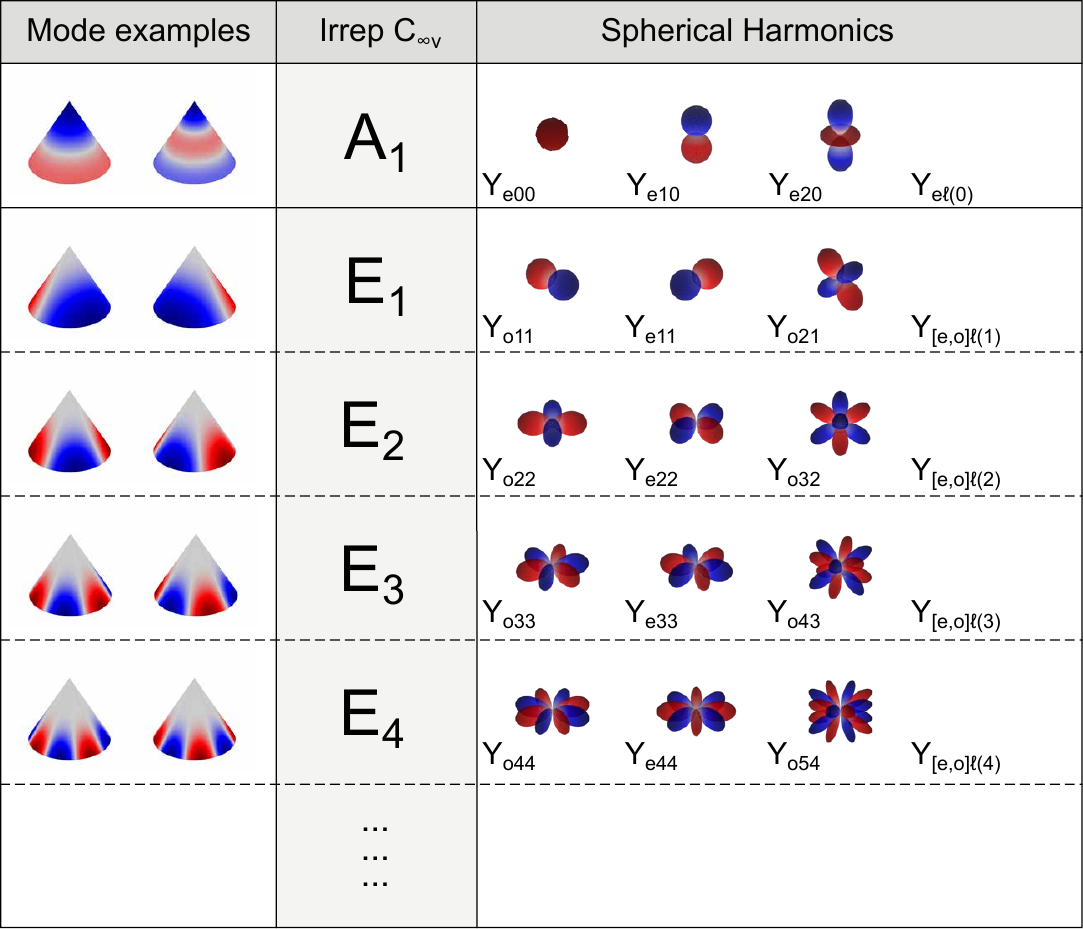}
        \end{tabular}
        &
        \vspace{0pt}
        \begin{tabular}[t]{r}
            \includegraphics[width = 0.49\textwidth]{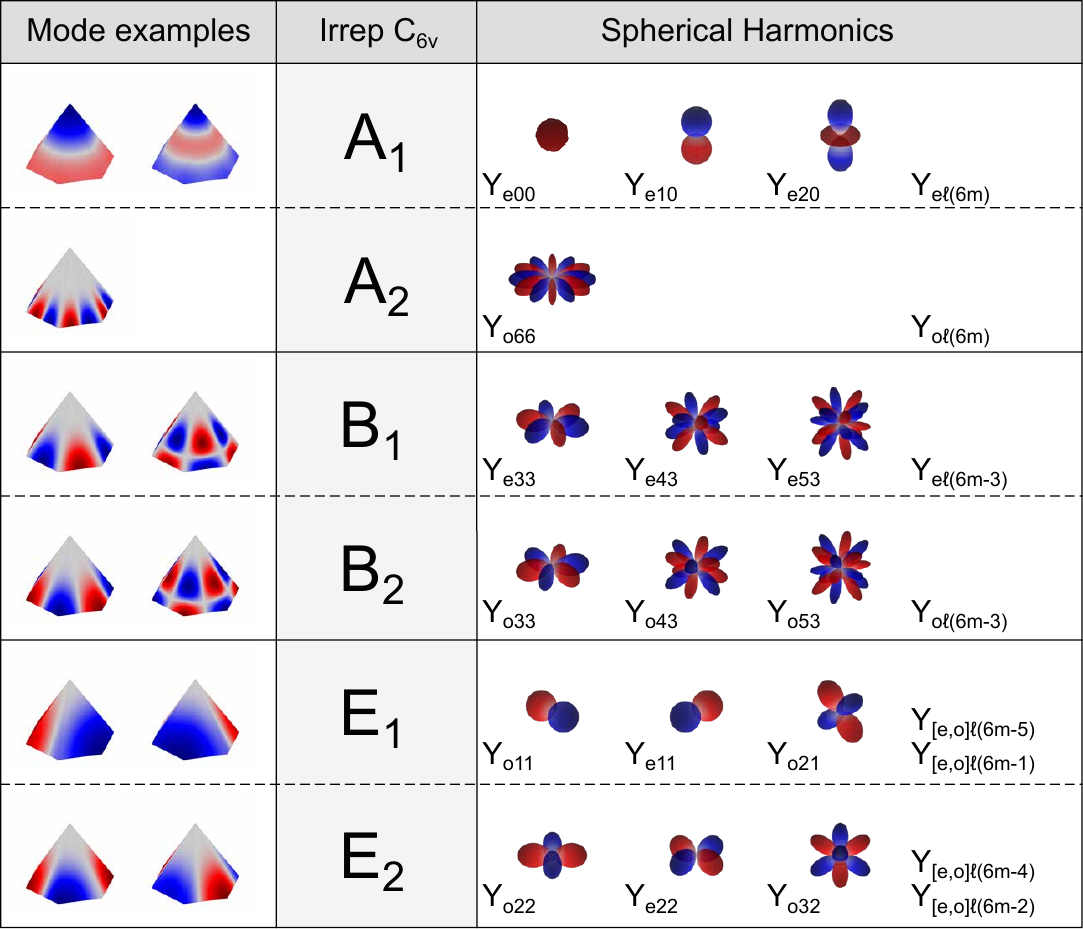}
            \end{tabular}
\\
        \vspace{0pt}
        \begin{tabular}[t]{c}
            \includegraphics[width = 0.49\textwidth]{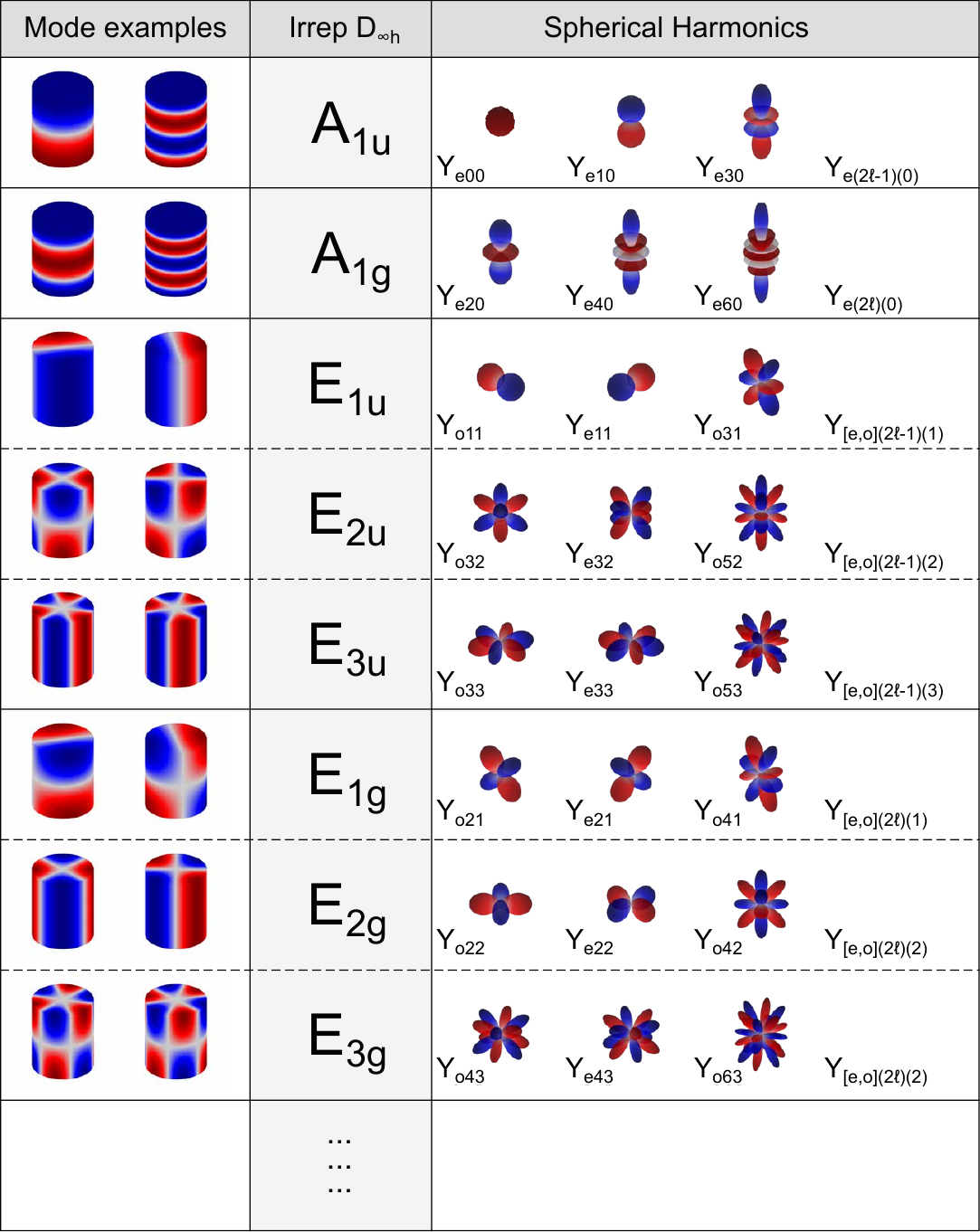}
        \end{tabular}  
        &
        \vspace{0pt}
        \begin{tabular}[t]{r}
            \includegraphics[width = 0.49\textwidth]{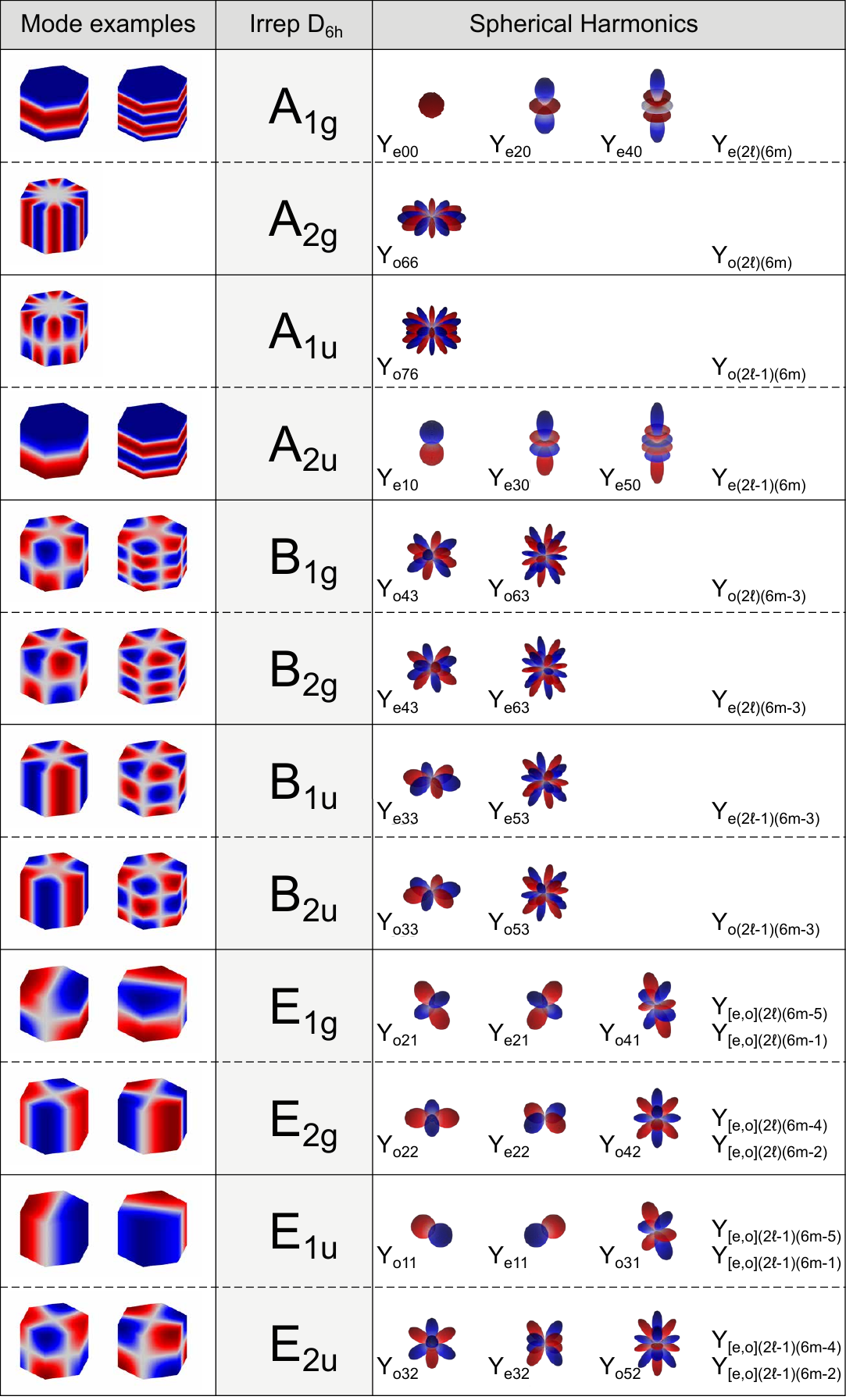}
        \end{tabular} 
\end{tabular}
\end{figure*}

\begin{figure*}[hp]
\caption{Tables of irreducible representations and examples of eigenmodes transformed under them, and tables of characters of irreducible representations for symmetry groups  $C_{2v}$, $C_{3v}$, $C_{4v}$}
\label{d4h}
\begin{tabular}{p{0.5\textwidth}p{0.5\textwidth}}
        \vspace{0pt}
        \begin{tabular}[t]{r}
            \includegraphics[width = 0.49\textwidth]{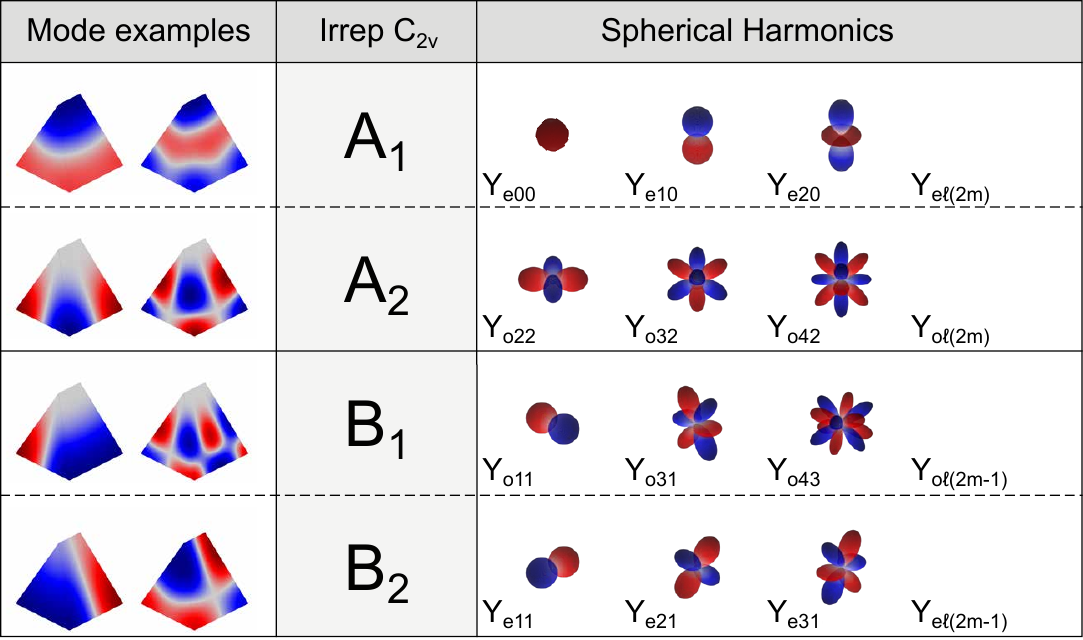}
            \end{tabular}
\\
        \vspace{0pt}
        \begin{tabular}[t]{c}
            \includegraphics[width = 0.49\textwidth]{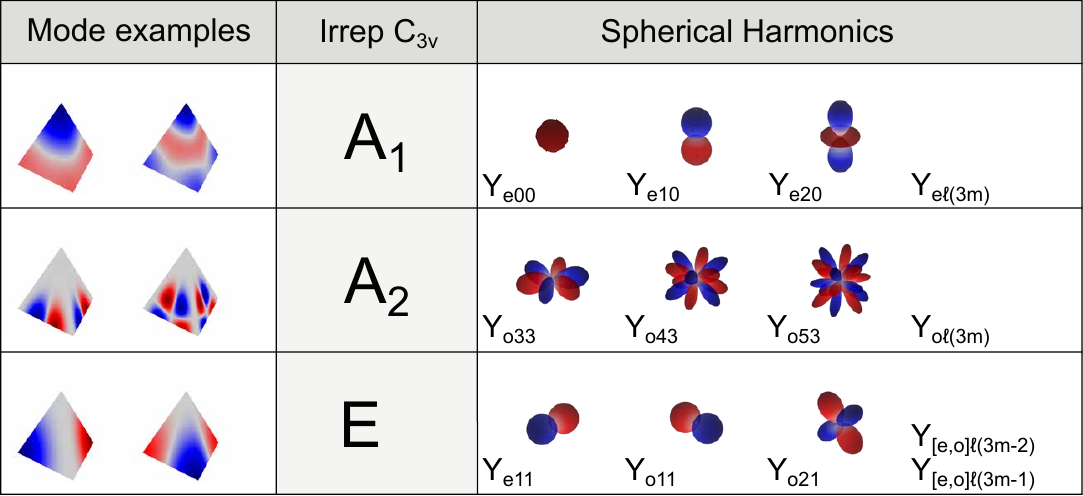}
        \end{tabular}  
\\
        \vspace{0pt}
        \begin{tabular}[t]{r}
            \includegraphics[width = 0.49\textwidth]{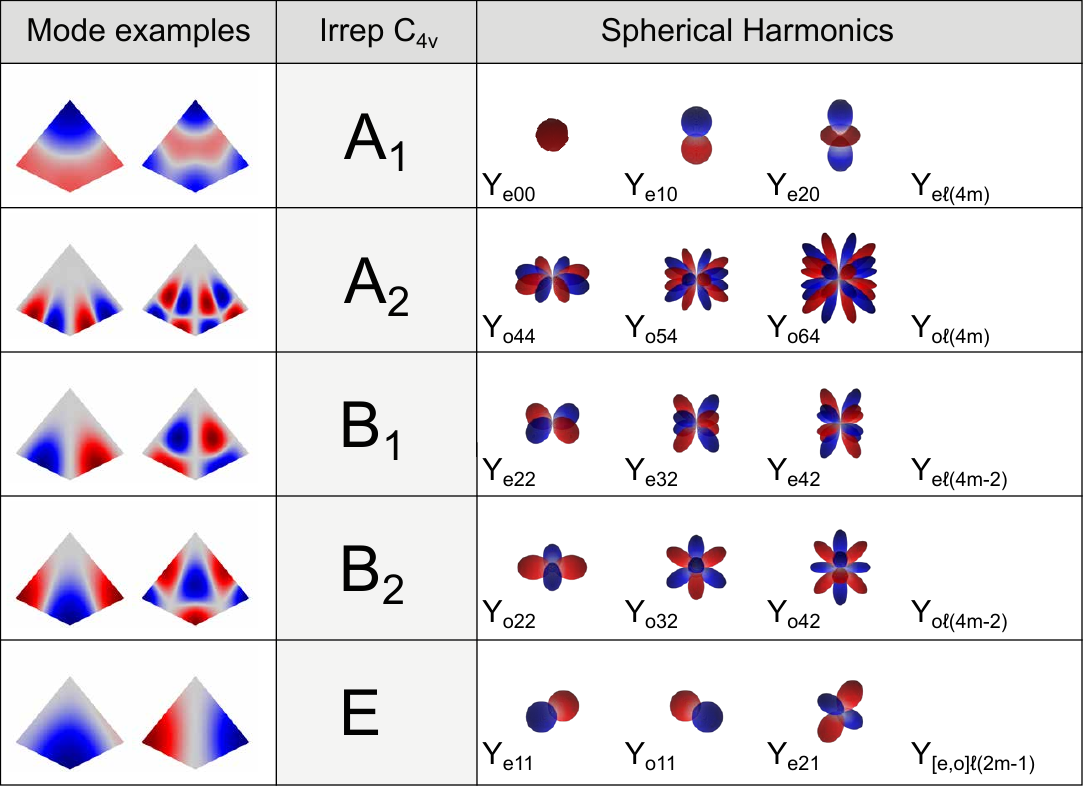}
        \end{tabular} 
\end{tabular}
\end{figure*}

\newpage 




\end{document}